\documentclass[aps,prd,amsmath,amssymb,reprint,superscriptaddress,nofootinbib]{revtex4-2}

\usepackage[colorlinks=true,allcolors=blue]{hyperref}
\usepackage{graphicx}\usepackage{dcolumn}\usepackage{bm}\usepackage[mathlines]{lineno}\usepackage{lipsum}
\usepackage{xspace}
\usepackage[nolist]{acronym}
\usepackage{color}
\usepackage[caption=false]{subfig}
\usepackage[inline]{enumitem}
\usepackage{bm}
\usepackage[normalem]{ulem}

\graphicspath{{images/}}

\newcommand\ifSI[1]{#1}
\newcommand\ifnotSI[1]{}
\newcommand\UnitSelect[2]{\ifnotSI{#1}\ifSI{#2}}

\usepackage{etoolbox}
\makeatletter
\patchcmd\linenumberpar{\@LN@parpgbrk}{\penalty\@LN@parpgpen\relax}{}{}
\makeatother

\makeatletter
\newcommand{\customlabel}[2]{\protected@write \@auxout {}{\string \newlabel {#1}{{#2}{\thepage}{#2}{#1}{}} }\hypertarget{#1}{#2}
}
\makeatother

\begin{document}

\title{Projected Sensitivity of DMRadio-m$^3$: A Search for the QCD Axion Below $1\,\mu$eV}

\author{L.~Brouwer}
\affiliation{Accelerator Technology and Applied Physics Division, Lawrence Berkeley National Laboratory, Berkeley, CA 94720}

\author{S.~Chaudhuri}
\affiliation{Department of Physics, Princeton University, Princeton, NJ 08544}

\author{H.-M.~Cho}
\affiliation{SLAC National Accelerator Laboratory, Menlo Park, CA 94025}

\author{J.~Corbin}
\affiliation{Department of Physics, Stanford University, Stanford, CA 94305}

\author{W.~Craddock}
\affiliation{SLAC National Accelerator Laboratory, Menlo Park, CA 94025}

\author{C.~S.~Dawson}
\affiliation{Department of Physics, Stanford University, Stanford, CA 94305}

\author{A.~Droster}
\affiliation{Department of Nuclear Engineering, University of California, Berkeley, Berkeley, CA 94720}

\author{J.~W.~Foster}
\affiliation{Center for Theoretical Physics, Massachusetts Institute of Technology, Cambridge, MA 02139}

\author{J.~T.~Fry}
\affiliation{Laboratory of Nuclear Science, Massachusetts Institute of Technology, Cambridge, MA 02139}

\author{P.~W.~Graham}
\affiliation{Department of Physics, Stanford University, Stanford, CA 94305}

\author{R.~Henning}
\affiliation{Department of Physics and Astronomy, University of North Carolina, Chapel Hill, Chapel Hill, North Carolina, 27599}
\affiliation{Triangle Universities Nuclear Laboratory, Durham, NC 27710}

\author{K.~D.~Irwin}
\email{irwin@stanford.edu}
\affiliation{Department of Physics, Stanford University, Stanford, CA 94305}
\affiliation{SLAC National Accelerator Laboratory, Menlo Park, CA 94025}

\author{F.~Kadribasic}
\affiliation{Department of Physics, Stanford University, Stanford, CA 94305}

\author{Y.~Kahn}
\affiliation{Department of Physics, University of Illinois at Urbana-Champaign, Urbana, IL 61801}

\author{A.~Keller}
\affiliation{Department of Nuclear Engineering, University of California, Berkeley, Berkeley, CA 94720}

\author{R.~Kolevatov}
\affiliation{Department of Physics, Princeton University, Princeton, NJ 08544}

\author{S.~Kuenstner}
\affiliation{Department of Physics, Stanford University, Stanford, CA 94305}

\author{A.~F.~Leder}
\affiliation{Department of Nuclear Engineering, University of California, Berkeley, Berkeley, CA 94720}
\affiliation{Accelerator Technology and Applied Physics Division, Lawrence Berkeley National Laboratory, Berkeley, CA 94720}

\author{D.~Li}
\affiliation{SLAC National Accelerator Laboratory, Menlo Park, CA 94025}

\author{J.~L.~Ouellet}
\email{ouelletj@mit.edu}
\affiliation{Laboratory of Nuclear Science, Massachusetts Institute of Technology, Cambridge, MA 02139}

\author{K.~M.~W.~Pappas}
\affiliation{Laboratory of Nuclear Science, Massachusetts Institute of Technology, Cambridge, MA 02139}

\author{A.~Phipps}
\affiliation{Department of Physics, California State University, East Bay, Hayward, CA 94542}
\author{N.~M.~Rapidis}
\affiliation{Department of Physics, Stanford University, Stanford, CA 94305}

\author{B.~R.~Safdi}
\affiliation{Department of Physics, University of California, Berkeley, Berkeley, CA 94720}

\author{C.~P.~Salemi}
\affiliation{Laboratory of Nuclear Science, Massachusetts Institute of Technology, Cambridge, MA 02139}

\author{M.~Simanovskaia}
\affiliation{Department of Physics, Stanford University, Stanford, CA 94305}

\author{J.~Singh}
\affiliation{Department of Physics, Stanford University, Stanford, CA 94305}

\author{E.~C.~van~Assendelft}
\affiliation{Department of Physics, Stanford University, Stanford, CA 94305}

\author{K.~van~Bibber}
\affiliation{Department of Nuclear Engineering, University of California, Berkeley, Berkeley, CA 94720}

\author{K.~Wells}
\affiliation{Department of Physics, Stanford University, Stanford, CA 94305}

\author{L.~Winslow}
\affiliation{Laboratory of Nuclear Science, Massachusetts Institute of Technology, Cambridge, MA 02139}

\author{W.~J.~Wisniewski}
\affiliation{SLAC National Accelerator Laboratory, Menlo Park, CA 94025}

\author{B.~A.~Young}
\affiliation{Department of Physics, Santa Clara University, Santa Clara, CA 95053}

\collaboration{DMRadio Collaboration}
\noaffiliation
 
\date{April 28, 2022}

\begin{abstract}
The QCD axion is one of the most compelling candidates to explain the dark matter abundance of the universe. With its extremely small mass ($\ll1\,\mathrm{eV}\ifSI{/c^2}$), axion dark matter interacts as a classical field rather than a particle. Its coupling to photons leads to a modification of Maxwell's equations that can be measured with extremely sensitive readout circuits. DMRadio-m$^3$ is a next-generation search for axion dark matter below $1\,\mu$eV using a $>4$\,T static magnetic field, a coaxial inductive pickup, a tunable LC resonator, and a DC-SQUID readout. It is designed to search for QCD axion dark matter over the range $20\,\mathrm{neV}\lesssim m_a\ifSI{c^2}\lesssim 800$\,neV ($5\,\mathrm{MHz}<\nu<200$\,MHz). The primary science goal aims to achieve DFSZ sensitivity above $m_a\ifSI{c^2}\approx120$\,neV (30\,MHz), with a secondary science goal of probing KSVZ axions down to  $m_a\ifSI{c^2}\approx40$\,neV (10\,MHz). \end{abstract}

\maketitle

\newcommand{\gagg}{\ensuremath{g_{a\gamma\gamma}}\xspace}
\newcommand{\gagggann}{\ensuremath{g_{a\gamma\gamma}g_{aNN}\xspace}}
\newcommand{\gagggaee}{\ensuremath{g_{a\gamma\gamma}g_{aee}\xspace}}

\newcommand{\rhoDM}{\ensuremath{\rho_{\rm DM}}\xspace}
\newcommand{\Jeff}{\ensuremath{\mathbf{J}_{\rm eff}}\xspace}

\newcommand{\cPU}{\ensuremath{c_{\rm PU}}\xspace}
\newcommand{\VPU}{\ensuremath{V_{\rm PU}}\xspace}
\newcommand{\LPU}{\ensuremath{L_{\rm PU}}\xspace}
\newcommand{\Leff}{\ensuremath{L_{\rm eff}}\xspace}

\newcommand{\DMR}{\mbox{DMRadio}\xspace}
\newcommand{\DMRp}{\mbox{\DMR-Pathfinder}\xspace}
\newcommand{\DMRL}{\mbox{\DMR-50L}\xspace}
\newcommand{\DMRm}{\mbox{\DMR-m$^3$}\xspace}
\newcommand{\DMRGUT}{\mbox{\DMR-GUT}\xspace}

\newcommand{\ABRA}{\mbox{ABRACADABRA}\xspace}
\newcommand{\ABRAten}{\mbox{ABRACADABRA-10\,cm}\xspace}

\begin{acronym}
\acro{SM}{Standard Model}
\acro{QED}{quantum electrodynamics}
\acro{QCD}{quantum chromodynamics}
\acro{BSM}{beyond the standard model}
\acro{DM}{dark matter}
\acro{CDM}{cold dark matter}
\acro{GUT}{grand unification theory}
\acro{WIMP}{weakly interacting massive particle}
\acro{SHM}{Standard Halo Model}
\acro{ppm}{part-per-million}
\acro{ppb}{part-per-billion}

\acro{ADM}{axion dark matter}
\acro{ALP}{axion-like particle}
\acro{PQ}{Peccei-Quinn}
\acro{PQWW}{Peccei-Quinn-Wilczek-Weinberg}
\acro{KSVZ}{Kim-Shifman–Vainshtein–Zakharov}
\acro{DFSZ}{Dine–Fischler–Srednicki–Zhitnitsky}
\acro{LSW}{light shining through wall}

\acro{DP}{dark photon}

\acro{DR}{dilution refrigerator}
\acro{PT}{pulse tube}
\acro{OFHC}{oxygen-free, high-conductivity}

\acro{TE}{transverse electric}
\acro{TM}{transverse magnetic}
\acro{TEM}{transverse electromagnetic}
\acro{MQS}{magneto-quasistatic}
\acro{SQL}{standard quantum limit}
\acro{QND}{quantum non-demolition}

\acro{DFT}{discrete Fourier transform}
\acro{FFT}{fast Fourier transform}
\acro{SNR}{signal-to-noise ratio}
\acro{PSD}{power spectral density}

\end{acronym}

\acrodefplural{PSD}{power spectral densities}
\acrodefplural{GUT}{grand unification theories}
\acrodefplural{ppm}{parts-per-million}
\acrodefplural{ppb}{parts-per-billion}

\section{Introduction}
\label{sec:Intro}

The Strong CP problem describes an unnaturally fine-tuned symmetry of nature that suggests an explanation beyond the \ac{SM} of particle physics. The leading solution to this problem is the introduction of a new \acl{PQ} symmetry, which is spontaneously broken at some high energy scale $f_a$ producing a pseudo-Goldstone boson, the axion $a$ \cite{Peccei:1977ur,Peccei:1977hh,Weinberg:1977ma,Wilczek:1977pj}. Interactions with \ac{QCD} give the axion a potential, which solves the Strong CP problem, gives the axion mass $m_a\ifSI{c^2}\approx5.7\,\mathrm{neV}(10^{15}\,\mathrm{GeV}/f_a)$ \cite{Borsanyi:2016ksw}, and produces a relic axion abundance in the early universe that satisfies the conditions to be \ac{DM} \cite{Abbott:1982af,Preskill1983,Dine1983}. Over the last few years, the axion has emerged as a leading \ac{DM} candidate. 

Recent theoretical work has opened a wide range of interesting parameter space for both \ac{QCD} axion and \ac{ALP} models \cite{Tegmark:2005dy,Hertzberg:2008wr,Co:2016xti,Graham:2018jyp,takahashi2018qcd}. In particular, the mass range $1\,\mathrm{neV}\lesssim m_a\ifSI{c^2} \lesssim 1\,\mu\mathrm{eV}$ is interesting for \acp{GUT} \cite{DiLuzio:2020wdo,PhysRevD.94.075001,Wise:1981ry,Ballesteros:2016xej,Ernst:2018bib,DiLuzio:2018gqe,Ernst:2018rod,FileviezPerez:2019fku,FileviezPerez:2019ssf,Co:2016xti}, String Theory models \cite{Svrcek:2006yi,Green:1984sg,Conlon:2006tq,Acharya:2010zx,Ringwald:2012cu,Cicoli:2012sz,Halverson:2019cmy,Witten:1984dg}, and naturalness arguments \cite{PhysRevD.73.023505,Graham:2018jyp}. Other \ac{ALP} models in this mass range simultaneously explain both the \ac{DM} abundance and matter-antimatter asymmetry \cite{Co2021rt}. 

Its small mass and cold temperature give \ac{ADM} a very high per-state occupation making it interact as a wave-like classical field. Depending on the precise model, the axion may couple to any \ac{SM} particle, but one of the least model-dependent couplings is the axion-photon coupling \gagg \cite{DiLuzio:2020wdo}. For the \ac{QCD} axion, \gagg is directly proportional to $m_a$, with an uncertainty on the proportionality constant spanning an order of magnitude in uncertainty and represented by the \ac{KSVZ} \cite{PhysRevLett.43.103,SHIFMAN1980493} and \ac{DFSZ} \cite{Dine:1981rt,Zhitnitsky:1980tq} models. A more generic class of \acp{ALP} break this proportionality and can have a wide range of masses and \gagg couplings.

The axion-photon coupling produces a modified Amp\`ere's law, behaving as an effective current that can be written approximately as
\begin{equation}
\Jeff\approx \gagg\ifSI{\frac{\sqrt{\hbar c}}{\mu_0}}\sqrt{2\rhoDM}\cos\left(\UnitSelect{m_at}{\frac{m_ac^2t}{\hbar}}\right)\mathbf{B},
\label{eqn:jeff}
\end{equation}
with local \ac{DM} density $\rhoDM\approx 0.45\,\mathrm{GeV/cm^3}$ \cite{de_Salas_2021} and axion frequency $\nu_a= m_a\ifSI{c^2}/(2\pi\ifSI{\hbar)}$. A powerful approach to searching for \ac{ADM} is to deploy a large, static \mbox{$\mathbf{B}$-field} to drive \Jeff through a pickup structure and search for excess power in a narrow axion signal bandwidth $\Delta \nu_{\rm sig}/\nu_a\approx10^{-6}$ set by the \ac{SHM} \cite{Herzog-Arbeitman:2017fte}.

\begin{figure*}[ht]
  \centering
  \subfloat[\label{fig:CoaxCartoon}]{\includegraphics[trim=50 50 50 50 ,clip,height=.29\textwidth]{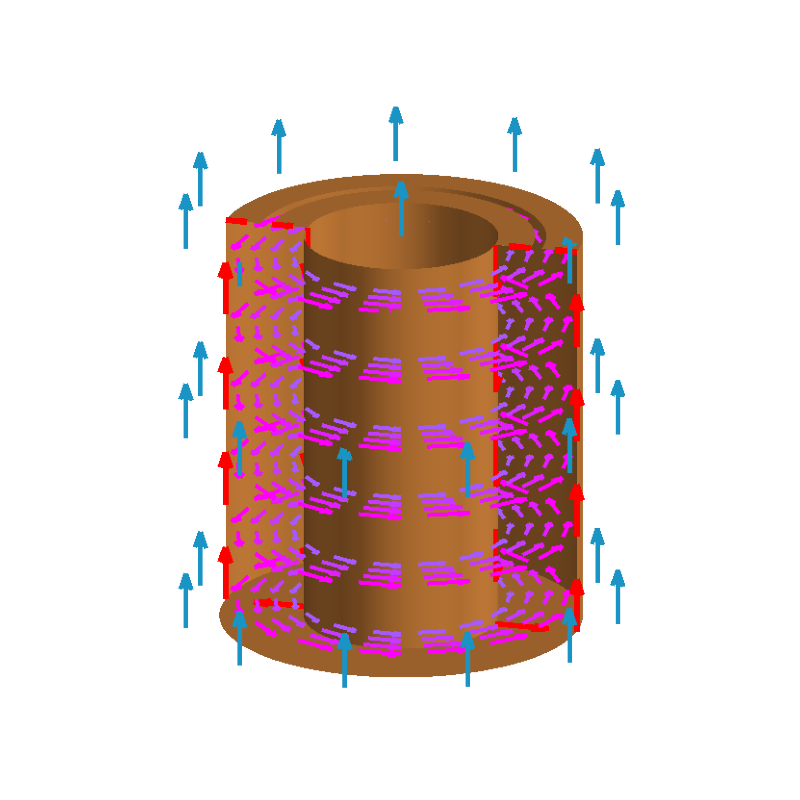}}\hspace{.6cm}
  \subfloat[\label{fig:DetectorDesign_CAD}]{\includegraphics[height=.29\textwidth]{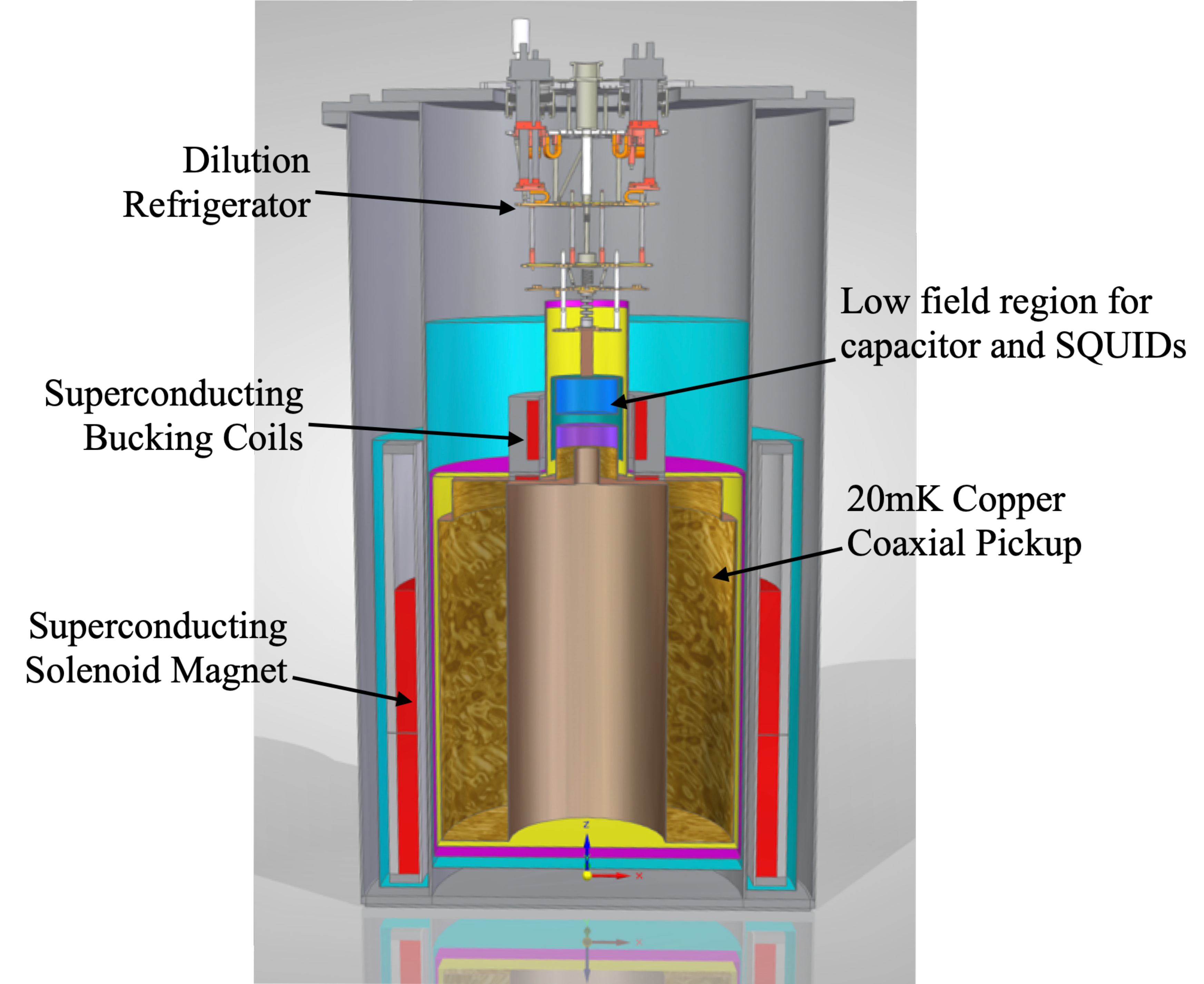}}\hspace{.6cm}
  \subfloat[\label{fig:DetectorDesign_MagnetProfile}]{\includegraphics[height=.29\textwidth]{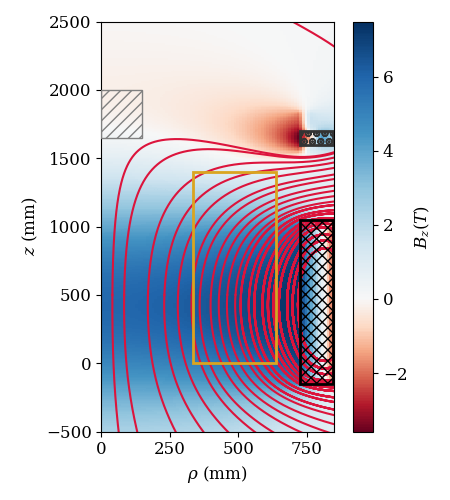}}
  \caption{Left: Cartoon of the \DMRm coaxial pickup in a uniform magnetic field. The magnet (not shown) drives an oscillating \Jeff (blue arrows) along the axis of the coax, which drives an azimuthal magnetic field (purple arrows), and a current along the coax (red arrows). Voltage accumulates across the slit on the top. Middle: Cross section of \DMRm with major components labeled. Total height is $\sim2.5$\,m. Right: Conceptual magnetic field design that produces high field across the volume of the coax (gold box) and a low field region for sensitive electronics (gray dashed region). Color shading represents the vertical component of the primary field $B_{z}$, while the red lines indicate the magnetic field flux lines. The field is generated by the solenoid coils in the black-hashed region, and bucking coils with counter flowing current in the black-dotted region.}
  \label{fig:DetectorDesign}
\end{figure*}

Historically, \ac{ADM} experiments have focused on the mass range $1\,\mu\mathrm{eV}\lesssim m_a\ifSI{c^2}\lesssim 40\,\mu\mathrm{eV}$, where the axion is able to resonantly excite a microwave cavity \cite{Hagmann1990,Asztalos2001,PhysRevLett.124.101303,PhysRevLett.127.261803,Brubaker2017,Backes2021,PhysRevD.103.102004,Lee2020}. However, the cavity resonance requires a detector size $L$ comparable to the axion Compton wavelength, $m_a \ifSI{c}L\ifSI{/\hbar} \sim1$ and makes probing masses below $1\,\mu\mathrm{eV}$ impractical. At lower axion masses, the \ac{MQS} limit applies and \Jeff can be treated as inducing magnetic fields that are detectable with an inductive pickup and enhanced with a lumped-element resonator \cite{Cabrera2008,PhysRevLett.112.131301,PhysRevLett.117.141801,PhysRevD.92.075012,DMRadio_Pathfinder}. This has been experimentally realized with a toroidal magnet with the \ABRAten prototype \cite{PhysRevLett.122.121802,PhysRevD.99.052012,PhysRevLett.127.081801} and SHAFT \cite{Gramolin2020a} and in a solenoidal geometry with ADMX-SLIC \cite{PhysRevLett.124.241101} and the BASE Penning trap \cite{PhysRevLett.126.041301}. \DMRp also performed a resonant search for \acp{DP} in the \ac{MQS} regime with a solenoidal pickup \cite{10.1007/978-3-030-43761-9_16}. In this letter, we present an optimized lumped-element experimental design called \DMRm, capable of probing \ac{ADM} over the mass range $20\,\mathrm{neV}\lesssim m_a\ifSI{c^2}\lesssim 800\,\mathrm{neV}$ ($5\,\mathrm{MHz}<\nu<200$\,MHz) with a 5\,yr scan time, achieving \ac{DFSZ} sensitivity for masses above $\sim\!120\,\mathrm{neV}\ifSI{/c^2}$ (30\,MHz).

\section{Lumped-Element Detection}
\label{sec:lumped_element_detectors}

The challenge facing any \ac{ADM} experiment is one of \ac{SNR}. The \ac{ADM} field contains enough power per square meter to illuminate an LED, but the self-impedance of photon-electron coupling implies that any practical receiver will only extract a tiny fraction of this power \cite{Chaudhuri2021dj}. Resonant circuits -- either microwave cavities or LC lumped-element circuits -- have much lower impedance on resonance and can extract more power from the axion field. 

For a readout circuit inductively coupled to \Jeff, the induced voltage can be expressed as 
\begin{equation}
|\tilde{V}_{pp}|^2 = 4\gagg^2\rhoDM\ifSI{\left(\frac{c^5}{\hbar \mu_0}\right)} m_a^2 \cPU^2 \Leff B_0^2 \VPU^{5/3}
\label{eqn:pickup_vpp}
\end{equation}
(see Appendix). Here $B_0$ and \VPU are the characteristic magnetic field and pickup volume, \Leff is the effective inductance of the pickup and readout circuit at the axion frequency $2\pi\nu_a=m_a\ifSI{/\hbar}$ and should ideally be dominated by the effective inductance of the pickup, \LPU. \cPU is a dimensionless proportionality constant that depends on the geometry. When coupled to a capacitor, the impedance of the resulting circuit $Z_L(\nu)$ decreases significantly at the resonance frequency, $\nu_r$. If $\nu_r$ is tuned close to the axion driving frequency, the current driven through the circuit from axion conversion, $\tilde{I}_{\rm sig}$, can be enhanced by multiple orders of magnitude  -- expressed in terms of the resonator quality $Q$:
\begin{equation}
|\tilde{I}_{\rm sig}|^2 =  4\gagg^2\rhoDM\ifSI{\left(\frac{\hbar c}{ \mu_0}\right)} \frac{\cPU^2 B_0^2 \VPU^{5/3}}{\Leff} 
\left(\frac{Q^2}{1+(2Q\Delta)^2}\right)
\label{eqn:sig_current}
\end{equation}
which holds for small fractional detunings ($\Delta\ll1$), and assumes that the pickup dominates the inductance of the circuit. The signal follows the Lorentzian response of the resonator circuit and depends on the fractional detuning of resonance from the axion frequency, \mbox{$\Delta = (2\pi\ifSI{\hbar}\nu_r-m_a\ifSI{c^2})/m_a\ifSI{c^2}$}. This gives the resonator characteristic width $\Delta\nu_r = \nu_r/Q$. 

The sensitivity for a resonant search is determined by its scan rate: the rate at which one can scan through frequencies searching for the axion induced signal current. The scan rate is set by the desired \gagg sensitivity and \ac{SNR} at each resonator tuning, and the total equivalent current noise $|\tilde{I}_N|^2$ at that frequency. At a single resonator tuning, the integration time, $\tau$, required to achieve this \ac{SNR} at \gagg is given by a modified Dicke radiometer equation
\begin{equation}
\mathrm{SNR}(\nu) = \frac{|\tilde{I}_{\rm sig}|^2}{|\tilde{I}_N(\nu)|^2}\sqrt{\tau\Delta\nu_{\rm sig}}.
\label{eqn:dicke_eqn}
\end{equation}
The current noise $|\tilde{I}_N(\nu)|^2$ is dominated by thermal Johnson-Nyquist noise and the noise of the first stage amplifier, $\eta_A(\nu)$. The amplifier noise consists of both the imprecision and backaction noise (see Appendix).

For a search with many discrete resonator tunings, the scan rate is approximated by the continuous function 
\begin{equation}
\begin{split}
\frac{d\log\nu_r}{dt} =  \pi&(6.4\times10^5) \ifSI{\left(\frac{c^2}{\mu_0^2}\right)} \frac{\gagg^4\rhoDM^2 }{\mathrm{SNR}^2\nu_r} \times\\ & c_{\rm PU}^4 QB_0^4\VPU^{10/3}\bar{\mathcal{G}}(\nu_r,\ifSI{k_B}T,\eta_A(\nu_r))
\end{split}
\label{eqn:scan_rate}
\end{equation}
$\bar{\mathcal{G}}(\nu_r,\ifSI{k_B}T,\eta_A(\nu_r))$ is a dimensionless factor that defines the effect of the amplifier matching on the scan rate (see Appendix). Explicitly, it is a trade-off between the Lorentzian signal gain and the frequency dependent system noise. Together, these two set the sensitivity bandwidth, $\Delta\nu_{\rm sens}$, which determines the spacing between adjacent resonator tunings. Since the thermal noise and amplifier backaction components follow the same Lorentzian lineshape as the signal gain, were they the only noise sources, we would have constant \ac{SNR} over all frequencies and $\Delta\nu_{\rm sens}$ could grow arbitrarily large. However, at frequencies sufficiently far from $\nu_r$, the amplifier imprecision noise dominates, decreasing the \ac{SNR}. For a scattering-mode circuit tuned for optimal single-frequency power transfer, the sensitivity bandwidth nearly matches the resonator line width, $\Delta\nu_{\rm sens}\approx \nu_r/Q$. Alternatively, for a matching circuit tuned for optimal scan speed, modestly reducing on-resonance \ac{SNR} for a larger $\Delta\nu_{\rm sens}$ can significantly improve the scan rate. In the thermal noise limit, the optimal coupling yields $\bar{\mathcal{G}}(\nu_r,\ifSI{k_B}T,\eta)\approx2\pi\ifSI{\hbar}\nu_r(6\sqrt{3}\ifSI{k_B}T\eta)^{-1}$.

\section{Detector Design}
\label{sec:detector_design}

\begin{figure}[ht]
  \centering
  \includegraphics[width=.48\textwidth]{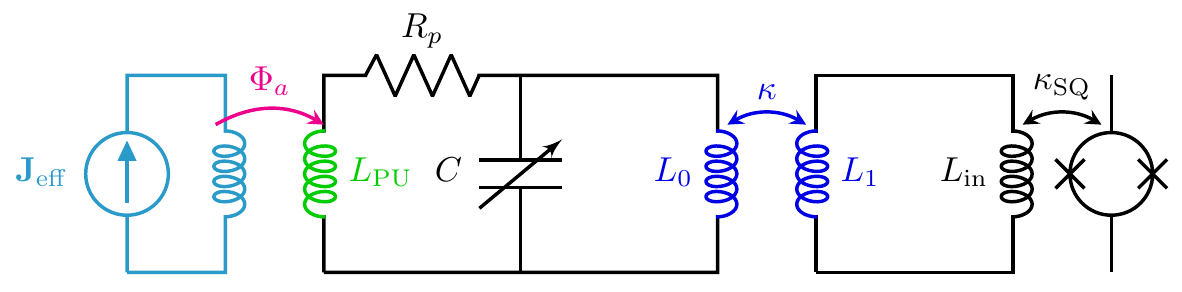}
  \caption{Effective \DMRm circuit model. \Jeff is inductively coupled to a coaxial pickup, $\LPU$, which, along with a tunable capacitor, $C$, form the resonant circuit. An inductive transformer $\kappa$ couples the signal to the SQUID amplifier. Loss in the readout is represented by resistor, $R_p$. On resonance, $\nu_{r} \approx 1/(2\pi \sqrt{\LPU C})$, the circuit impedance drops and axion power is more efficiently transferred to the receiver.}
  \label{fig:circuit_diagram}
\end{figure}

\begin{figure*}[ht]
    \centering
    \includegraphics[width=.94\textwidth]{DMRadio_Sensitivity_Projections\ifSI{_SI}_labels_v2.pdf}
    \caption{Projected \DMRm \ac{ADM} sensitivity with a 5 year scan, with projected sensitivity for \DMRL, and existing bounds below $1\,\mu$eV from \ABRAten, SHAFT, \mbox{ADMX-SLIC} (spike around 200\,neV), BASE (spike around 3\,neV), and bounds from microwave cavities above $1\,\mu$eV. \DMRm targets DFSZ sensitivity above $\approx120\,$neV and secondarily KSVZ above $\approx40\,$neV. Plot generated using code from \cite{OHareAxionLimits}.}
    \label{fig:sensitivity}
\end{figure*}

\DMRm uses a static $>$4\,T solenoid magnet to drive \Jeff along the axis of a coaxial inductive pickup (see Fig.~\ref{fig:DetectorDesign_CAD}.) The coaxial pickup couples to a tunable capacitor to form an LC resonator circuit, targeting a quality factor of $Q>10^5$. The circuit is read out by conventional DC-SQUIDs with a target noise level of $20\times$ the \ac{SQL}. The coaxial pickup and capacitor are cooled to an operating temperature of $T\approx20$\,mK, while the magnet sits at 4\,K. We discuss each of these in detail below.

\DMRm builds on the experience of \ABRAten, \DMRp, and \DMRL (currently under construction.) All three detectors use a superconducting inductive pickup to search for \ac{ADM} or \acp{DP} below 20\,neV. \DMRp and \DMRL \cite{DMRadio-50L} utilize a resonant readout approach, and \ABRAten and \DMRL utilize a toroidal magnet. The toroidal geometry works well at low frequencies (i.e.\ low axion masses) because it completely encloses the magnetic field, allowing the use of lossless superconducting materials. Loss is only introduced through coupling to lossy materials near the detector, which can be efficiently screened. However, at frequencies above $\sim$50\,MHz, irreducible capacitances associated with the readout begin to short the signal and rapidly reduce the signal sensitivity at higher frequencies \cite{ToroidVsSolenoid}. A change of geometry is required to probe higher axion masses.

\DMRm uses a solenoid magnet, with a bore volume of $V\approx 2\,\mathrm{m}^3$, and a characteristic field of $B_0 >4\,\mathrm{T}$. This geometry creates a large-volume, high-field region and drives \Jeff along the axis of the solenoid (see Fig.~\ref{fig:CoaxCartoon}.) Unlike the toroid, the solenoid exposes the magnetic field, requiring that the pickup sit in the large primary field. Because SQUIDs cannot operate in such a field, \DMRm uses a set of bucking coils to bend the field lines outward, producing a low-field region directly above the high-field region. The resulting magnetic field has significant variation within the bore of the magnet, reaching a peak value of $\approx7\,\mathrm{T}$ off-axis; but with a characteristic value of $B_0 > 4\,\mathrm{T}$ along the central axis in the high-field region. The low-field region sits approximately 20\,cm above the high-field region and has a maximum field of $|\mathbf{B}|\lesssim400\,\mathrm{mT}$ (see Fig.~\ref{fig:DetectorDesign_MagnetProfile}). These remaining fields are further reduced with additional bucking coils and superconducting shielding.

\DMRm detects the axion-induced \Jeff via a coaxial inductive pickup in the high-field region of the magnet (see Fig.~\ref{fig:CoaxCartoon}). The geometry maximizes the coupling to \Jeff, while minimizing the pickup inductance \LPU. Because it sits inside the magnetic field, it is challenging to make the pickup from lossless superconducting material. Instead, the pickup is made from \ac{OFHC} copper. The finite conductivity of copper sets an upper limit on the achievable quality factor of the LC resonator. At \DMRm frequencies and at temperatures $<100\,\mathrm{mK}$, the anomalous surface resistance of copper is lower than at $\sim$GHz frequencies and, combined with the large volume-to-surface ratio, the resonator goal of $Q>10^5$ is achievable.  

While the inductive pickup can sit in the primary field without unacceptably degrading $Q$, the same does not extend to the tunable capacitor, which must be superconducting, or to the SQUIDs, which require an ultra-low field to operate. These are instead placed in the low-field region above the coaxial pickup. To bridge the distance between the pickup and the low-field region without introducing excessive parasitic inductance, the coax funnels to a narrow neck coupled to the matching circuit above. To probe the mass range from \mbox{$20\,\mathrm{neV}\lesssim m_a\ifSI{c^2}\lesssim 800\,\mathrm{neV}$}, \DMRm uses a set of interchangeable coaxial pickups that are exchanged during the run of the experiment. The coaxial pickups are currently being designed, and aim to keep $\cPU$ \& $\VPU$ maximized over the entire search range, while avoiding resonant modes that may short our signal.

The matching circuit must optimally couple to the pickup and convey the signal to the SQUID readout (see Fig.~\ref{fig:circuit_diagram}) and consists of the coupling transformer, a tunable capacitor and the inductive coupling to the SQUID amplifier. Any additional inductance in the circuit that is not directly coupled to the signal flux  reduces the coupled power. In the series transformer configuration of \cite{PhysRevLett.112.131301,PhysRevLett.117.141801}, the coupled axion power is maximized by minimizing the transformer inductance $L_0\ll L_{\rm PU}$, which is technically challenging given the nominal $\LPU\lesssim200\,\mathrm{nH}$. However, in the parallel transformer configuration of Fig.~\ref{fig:circuit_diagram}, axion signal power is maximized by making the transformer inductance $L_0\gg L_{\rm PU}$, which can be achieved more readily, with a transformer inductance of $L_0\gtrsim1\,\mu\mathrm{H}$.

The baseline tunable capacitor consists of an array of parallel superconducting plates and sapphire dielectrics. The tuning must be able to scan the range $10\,\mathrm{pF} \lesssim C \lesssim 5\,\mathrm{nF}$ with \ac{ppm} stepping precision. This can be achieved using a multiple capacitor design with a coarse tuning to give large capacitance swing to cover the full frequency range, and fine tuning to give \ac{ppm} frequency precision. If the loss in the insertable sapphire dielectric proves to be too large, degrading the resonator $Q$ to an unacceptable level, a fallback option includes using movable superconducting capacitor plates alone to adjust conductor overlap and resultant capacitance. The design of the tunable capacitor will build off the one currently under construction for \DMRL where many of these questions are being addressed. 

Finally, the signal is amplified and read out through a phase-insensitive DC SQUID readout. To maximize the frequency-integrated axion sensitivity, the coupling transformer coefficient $\kappa$ must maintain an optimal tradeoff between imprecision and backaction noise. This optimization is worked out for a simple series circuit in detail in \cite{Chaudhuri:2018rqn}. For the sensitivity projections presented here, we target an amplifier noise level of $\eta_A=20$, i.e.\ $20\times$ the \ac{SQL}, which has been demonstrated in \cite{doi:10.1063/1.3002321}.

The optimal coupling $\kappa$ is in general frequency dependent, reflecting the fact that at $T\approx20\,\mathrm{mK}$ the number of thermal noise photons \mbox{$n_T=(\mathrm{exp}(2\pi\ifSI{\hbar}\nu/k_BT)-1)^{-1}$} depends strongly on frequency, and varies from $n_T\approx90$ at 5\,MHz to $n_T\approx2$ at 200\,MHz. Therefore, DMRadio-m$^3$ utilizes a transformer capable of \textit{in situ} variation of the coupling to maintain optimal matching to the readout amplifier.

The entire pickup structure is cooled to $T\approx20\,\mathrm{mK}$ using a \ac{DR}, while the magnet is held at 4\,K.  \acp{DR} cooling payloads of similar heat capacity and volume to lower temperatures have been implemented previously \cite{CUORECryostat,CUOREOneTon}. \DMRm will be located in Building 750 at SLAC National Accelerator Laboratory, the former site of the SLD detector.

\section{Sensitivity}
\label{sec:sensitivity}

\DMRm will scan the full range $20\,\mathrm{neV}\lesssim m_a\ifSI{c^2} \lesssim 800\,\mathrm{neV}$ in stages, by sequentially stepping through a series of detector configurations, each with a different coaxial pickup. For the lowest frequency coax $\cPU \approx 0.16$, with the exact value depending on  details of the final design.

The projected sensitivity for \DMRm is shown in Fig.~\ref{fig:sensitivity} alongside the sensitivity of the upcoming \DMRL, which will the subject of future publications. The shape of the sensitivity reach is set by the scan strategy, which aims to cover the full mass range with as much \ac{DFSZ} sensitivity as achievable in 5\,yr of scan time, leading to the corner at $\nu\approx30$\,MHz or $m_a\ifSI{c^2}\approx120\,\mathrm{neV}$. 

An intriguing feature of the sensitivity is that much of the scan time is spent at lower masses above the \ac{DFSZ} line. Above the 120\,neV corner, the target \gagg sensitivity increases $\propto \nu$, and the scan speed increases $\propto \nu^3$ (see Eq.~\ref{eqn:scan_rate}). The scan rate will be further enhanced because of the growth of $\bar{\mathcal{G}}$, due to the decreasing impact of thermal noise at higher frequencies -- though this growth is slightly slower than $\propto \nu_r$ and will be offset by a gradual decrease in SQUID sensitivity at higher frequency. 
 
\section{Conclusion}
\label{sec:conclusion}

The axion is among the best motivated candidates to explain the \ac{DM} abundance of the universe. At present, the parameter space of \ac{ADM} models is relatively unexplored. Recent advances in quantum sensors, magnet technology, and cryogenics have made searching the full \ac{ADM} parameter space more feasible than ever. The region of \ac{ADM} parameter space with $m_a\lesssim1\,\mu$eV is well motivated and particularly attractive due to its connection with \ac{GUT} models.

In this letter, we have presented the baseline design and sensitivity reach of the \DMRm experiment, which uses a lumped-element approach to search for \ac{QCD} \ac{ADM} over mass range $20\,\mathrm{neV}\lesssim m_a\ifSI{c^2} \lesssim 800\,\mathrm{neV}$. This design is capable of probing at or below the \ac{KSVZ} sensitivity over the full range and \ac{DFSZ} sensitivity above 120\,neV with a 5\,yr scan time. This will make \DMRm a key component of the next generation of experimental searches for \ac{ADM}. 

Future upgrades to \DMRm could moderately extend the \ac{DFSZ} sensitivity below the 120\,neV corner through the use of backaction-evading quantum sensors \cite{clerk2008back}. Future proposed experiments like \DMRGUT could probe \ac{DFSZ} models to even lower masses through the use of beyond-\ac{SQL} sensors and high-field, high-volume magnets \cite{Brouwer:2022bwo}.

\begin{acknowledgments}
The authors acknowledge support for \DMRm as part of the DOE Dark Matter New Initiatives program under SLAC FWP 100559. Members of the \DMR Collaboration acknowledge support from the NSF under awards 2110720 and 2014215. S.~Chaudhuri acknowledges support from the R.~H.~Dicke Postdoctoral Fellowship and Dave Wilkinson Fund at Princeton University. C.~P.~Salemi is supported in part by the National Science Foundation Graduate Research Fellowship under Grant No.~1122374.  Y.~Kahn was supported in part by DOE grant DE-SC0015655.  B.~R.~Safdi  was  supported  in part  by  the  DOE  Early  Career  Grant  DESC0019225. P.~W.~Graham acknowledges support from the Simons Investigator Award no. 824870 and the Gordon and Betty Moore Foundation Grant no. 7946. J.~W.~Foster was supported by a Pappalardo Fellowship.
 \end{acknowledgments}

\appendix
\section{Appendix: Scan Rate Calculation}
\label{sec:ScanRateCalc}

The general derivation of Eq.~\ref{eqn:scan_rate} and $\bar{\mathcal{G}}$ can be found in \cite{Chaudhuri:2018rqn}. Here we summarize the result for convenience.

In general, the voltage across an inductive pickup, \Leff, coupled to \Jeff will be proportional to the coupled energy $U_{\rm DM}$ \cite{Chaudhuri:2018rqn}
\begin{equation}
|V_{pp}^2| = 4 \ifSI{\left(\frac{c^4}{\hbar^2}\right)} m_a^2\Leff U_{\rm DM}
\label{eqn:vpp_Udm}
\end{equation}
The coupled energy, defined as
\begin{equation}
U_{\rm DM} = k^2\frac{\gagg^2 \rhoDM}{m_a^2}\ifSI{\frac{\hbar^3}{c\mu_0}}\int |\mathbf{B}(\mathbf{x})|^2 dV,
\label{eqn:Udm}
\end{equation}
gives a measure of the amount of energy coupled from the axion field into the pickup. The volume integral is performed over all space and is proportional to the total stored energy of the magnet. The dimensionless proportionality constant, $k$, is a geometric factor containing all the information about mutual coupling between the inductor and \Jeff. This constant generalizes the $C_{nml}$ form factor in a cavity haloscope. Energy conservation dictates that $k^2\leq\frac12$ \cite{Chaudhuri2021dj}. However, it is also clear that the voltage driven across an inductive element should go to zero as the driving frequency goes to zero ($m_a\rightarrow0$), and that therefore $k$ should have a dependence on $m_a$. Given this, it is convenient to define a new dimensionless constant, \cPU, that separates this mass dependence
\begin{equation}
k^2 = \ifSI{\left(\frac{c}{\hbar}\right)^2}m_a^2\VPU^{2/3}\left(\frac{B_0^2\VPU}{\int |\mathbf{B}(\mathbf{x})|^2\,dV}\right)\cPU^2
\label{eqn:cpu_def}
\end{equation}
and is both frequency and scale invariant. This constant can be calculated explicitly in the \ac{MQS} limit, when the Compton wavelength is large compared to the size of the detector. In this limit, we can use the relationship $V = \partial\Phi/\partial t$ and write $\tilde{V}_{pp} = 2\pi\nu_ai\tilde{\Phi}_a$ in the frequency domain, where $\Phi_a$ is the axion induced flux through the pickup. This flux can be extracted from the Biot-Savart law as
\begin{equation}
\begin{split}
\tilde{\Phi}_a = {} & \frac{\gagg\ifSI{\sqrt{\hbar c}}\sqrt{2\rhoDM}}{4\pi} \\
&\times\int_A dA\int dV'\mathbf{\hat{n}}\cdot \frac{\mathbf{B}_0(\mathbf{r}')\times(\mathbf{r}-\mathbf{r}')}{|\mathbf{r}-\mathbf{r}'|^3}.
\end{split}
\label{eqn:mqs_flux}
\end{equation}
where the integral $dA$ is taken over the area of the pickup $A$, and the integral $dV'$ is performed over the shielding volume containing the coax. Combining Eqns.~\ref{eqn:vpp_Udm}, \ref{eqn:Udm}, \ref{eqn:cpu_def}, and \ref{eqn:mqs_flux}, we can write \cPU in the \ac{MQS} limit as
\begin{equation}
\cPU^2 = \ifSI{\mu_0}\frac{\left|\int\,dA \int\,dV' \mathbf{\hat{n}}\cdot\frac{\mathbf{B}_0(\mathbf{r}')\times\mathbf{(\mathbf{r}-\mathbf{r}')}}{|\mathbf{r}-\mathbf{r}'|^3}\right|^2}{2 (4\pi)^{2}  B_0^2\VPU^{5/3}\Leff }.
\label{eqn:cpu_mqs}
\end{equation}
The scale invariance can be seen explicitly by noting the scaling $\Leff\approx\LPU\propto V^{1/3}$ for a constant geometry. Typical values for the \DMRm geometry are $\cPU\approx0.1-0.2$.

This formulation also generalizes beyond the validity of the lumped element approximation. At frequencies where the Compton wavelength becomes comparable to the size of the detector, $2\pi\ifSI{\hbar}/m_a\ifSI{c} \approx \VPU^{1/3}$, the coaxial pickup no longer behaves purely inductively and we must incorporate its full complex impedance into the calculation: $2\pi i\nu\LPU\rightarrow Z_{\rm PU}(\nu)$. We can Taylor expand any resonant circuit's impedance around the resonant frequency to find an effective inductance, 
\begin{equation}
Z(\nu) \approx R_{\rm eff}+4\pi i\Leff(\nu-\nu_r)+O\left((\nu-\nu_r)^2\right)
\end{equation}
This expansion is generally valid over frequency bandwidths small compared to the resonance frequency, i.e. for small fractional detunings $\Delta  \ll 1$\footnote{A more complete discussion of lumped elements detectors beyond the validity of the \ac{MQS} approximation is in preparation \cite{MQS_Breakdown}.}.

The resonant enhancement of the circuit comes directly from its impedance $Z(\nu)$
\begin{equation}
|\tilde{I}_{\rm sig}|^2=\frac{|\tilde{V}_{pp}|^2}{|Z(\nu)|^2}.
\end{equation}
For a generic RLC circuit, $Z(\nu) = R+i(2\pi \nu L +(2\pi \nu C)^{-1})$, we can evaluate
\begin{equation}
\begin{split}
\frac{1}{|Z(\nu)|^2} & {} = \frac{Q^2}{4\pi^2\nu_r^2 L^2\left(1+Q^2\frac{\nu_r^2}{\nu^2}\left(\frac{\nu^2}{\nu_r^2}-1\right)^2\right)}\\
& \approx\frac{Q^2}{4\pi^2\nu_r^2L^2 \left(1+4Q^2\Delta^2\right)} \\
\end{split}
\end{equation}
where $\nu_r=(2\pi\sqrt{LC})^{-1}$ and $Q=\sqrt{L/CR^2}$, and the approximation holds for $\nu\approx\nu_r$. We see the Lorentzian behavior of the resonator stems directly from the impedance of the readout circuit.

\begin{figure}[ht]
  \centering
  \includegraphics[width=.45\textwidth]{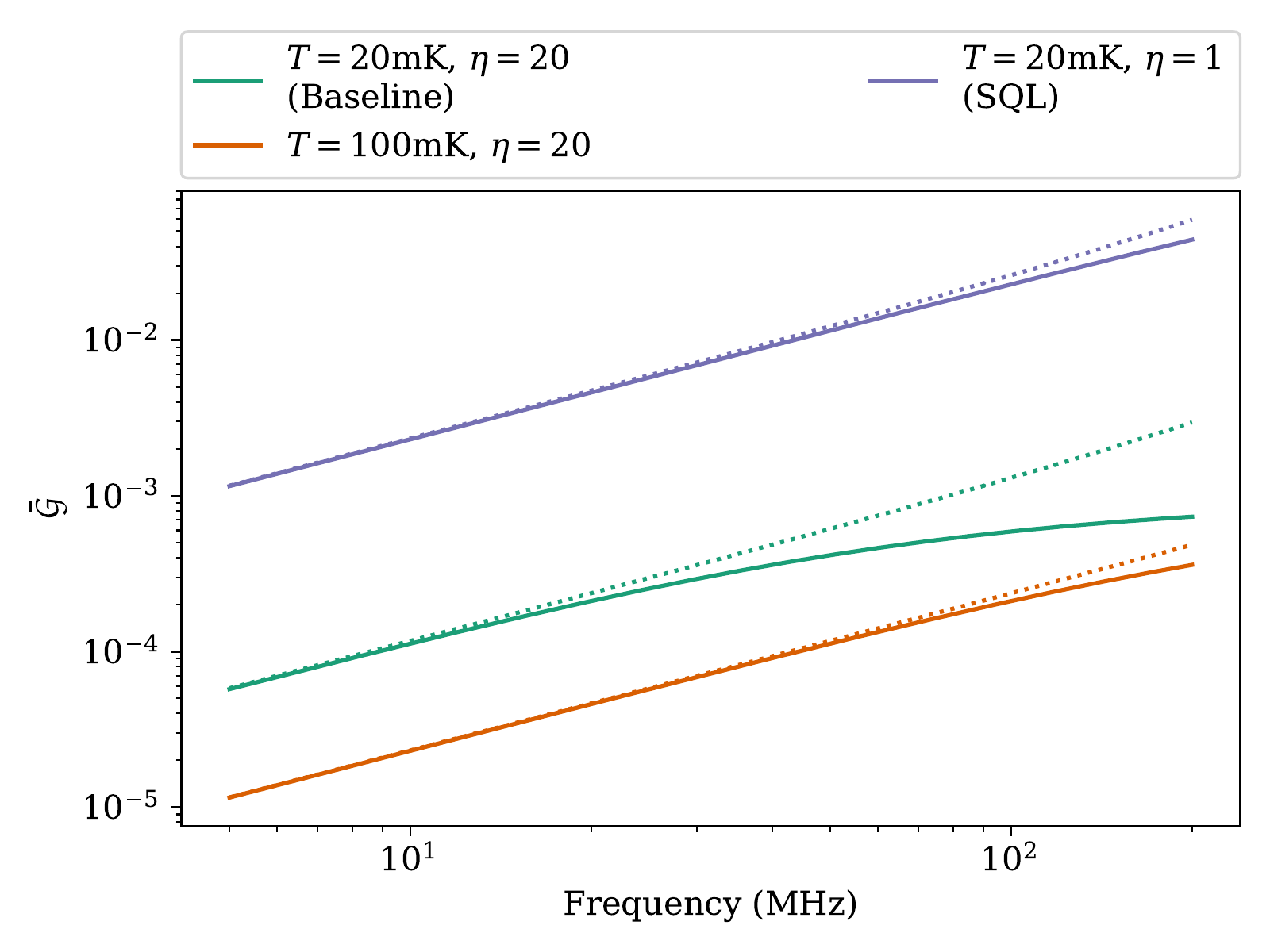}
  \caption{$\bar{\mathcal{G}}(\nu,\ifSI{k_B}T,\eta_A)$ as a function of frequency for several choices of physical temperature, $T$, and amplifier noise $\eta_A$. The solid lines show the full evaluation, while the dotted lines show the linear behavior, which holds for the region where $\ifSI{k_B}T/\ifSI{\hbar}\nu\gg\frac12\eta_A,1$.}
  \label{fig:squiggly_G}
\end{figure}

The noise power $|\tilde{I}_N(\nu)|^2$ can be expressed in terms of a current noise power spectral density at the input coil of the readout SQUID, 
\begin{equation}
S_{II}^{\rm tot}(\nu) = \frac{|\tilde{I}_N(\nu)|^2}{\Delta\nu}.
\end{equation} 
This noise power can be decomposed into thermal, amplifier, and vacuum components
\begin{equation}
\begin{split}
S_{II}^{\rm tot}(\nu)  = {} & 8\pi\ifSI{\hbar}\nu\left(n_T+\frac12\right)\frac{\mathrm{Re}[Z(\nu)]}{|Z(\nu)|^2}\\
& + S_{II}^{\rm Imp} + \frac{S_{VV}^{\rm BA}}{|Z(\nu)|^2}\end{split}
\label{eqn:noise_sources}
\end{equation}
Where the first term in the parenthesis corresponds to the number of thermal noise photons \mbox{$n_T(\nu,\ifSI{k_B}T)=\left(\exp(2\pi\ifSI{\hbar}\nu/\ifSI{k_B}T)-1\right)^{-1}$} at physical temperature $T$. \DMRm targets a physical temperature of $T = 20$\,mK, corresponding to $n_T\approx90$ noise photons at 5\,MHz and $n_T\approx2$ noise photons at 200\,MHz. The $\frac12$ corresponds to vacuum noise, intrinsic to any phase-insensitive measurement. $S_{II}^{\rm Imp}$ is the amplifier imprecision noise, expressed in terms of current noise at the amplifier input, and $S_{VV}^{\rm BA}$ is the amplifier backaction noise, expressed as a voltage source at the amplifier input and are both independent of frequency in this form.  Equation~\ref{eqn:noise_sources} demonstrates the different spectral behavior of the various noise sources. The thermal, vacuum, and backaction noise are all shaped by the resonator Lorentzian, while the imprecision noise has a spectrally flat distribution. In practice, a DC-SQUID will have an additional type of noise term, $S_{IV}^{\rm corr}$, that defines correlations between the imprecision and backaction noise \cite{MYERS2007182,RevModPhys.82.1155}. These correlations complicate the optimization and are included in our sensitivity calculation, but do not qualitatively change the approach and so are omitted here for simplicity.

The amplifier noise parameter, $\eta_A$ can be written in terms of these noise powers as
\begin{equation}
\eta_A = \frac{\sqrt{S_{II}^{\rm Imp}S_{VV}^{\rm BA}}}{2\pi\ifSI{\hbar}\nu}.
\end{equation}\\
At the \ac{SQL}, $\eta_A=1$. For \DMRm, the amplifier noise target is $\eta_A=20$, including the correlated noise that we have omitted here. This corresponds to $n_A=\eta_A/2=10$ added noise photons. 

The amplifier noise parameter $\eta_A$ depends on the amplifier imprecision and backaction noise as well as $\kappa$ coupling in Fig.~\ref{fig:circuit_diagram}, which regulates the tradeoff between the two. In particular, $S_{II}^{\rm Imp}\propto\eta_A/\kappa^2$, while $S_{VV}^{BA}\propto\eta_A\kappa^2$. A stronger coupling increases the backaction noise and decreases the input referred imprecision noise. Because of the different spectral responses of these two noise terms, the coupling $\kappa$ can be optimized to give the fastest possible scan rate. For a careful and extensive analysis of this optimization, including noise correlations, see \cite{Chaudhuri:2018rqn}. But the resulting effect on the scan rate can be written in terms of a single parameter 
\begin{equation}
\bar{\mathcal{G}} \left(\nu_r,\bar\alpha,\ifSI{k_B}T,\eta_A\right) = \frac{\bar\alpha}{\left[\bar\alpha^2 + 2(2n_T+1)\bar\alpha+\eta_A^2 \right]^{3/2}}
\end{equation}
where
\begin{equation}
\bar\alpha = \frac{2\eta_A^2}{2n_T + 1 + \sqrt{(2n_T+1)^2+8\eta_A^2}}.
\end{equation}
In the limit where $n_T\gg\frac12\eta_A,1$, or equivalently $\ifSI{k_B}T/\ifSI{\hbar}\nu\gg\frac12\eta_A,1$, $\bar{\mathcal{G}}$ has an approximately linear dependence on $\nu$:
\begin{equation}
\bar{\mathcal{G}} \approx \frac{2\pi\ifSI{\hbar}\nu}{6\sqrt{3}\ifSI{k_B}T\eta_A},\hspace{7mm}\frac{\ifSI{k_B}T}{\ifSI{\hbar}\nu}\gg\frac12\eta_A,1.
\end{equation}

\bibliographystyle{apsrev4-2}

\end{document}